# Multi-terminal memristive devices enabling tunable synaptic plasticity in neuromorphic hardware: a mini-review

**Yann Beilliard[1,2*], Fabien Alibart[2,3]**

[1]Institut Interdisciplinaire d'Innovation Technologique (3IT), Université de Sherbrooke, 3000, boulevard de l'Université, Sherbrooke (Québec) J1K 0A5, Canada

[2]Laboratoire Nanotechnologies Nanosystèmes (LN2) – CNRS, Université de Sherbrooke – 3000, boulevard de l'Université, Sherbrooke (Québec) J1K 0A5, Canada

[3]Institute of Electronics, Microelectronics and Nanotechnology (IEMN), Université de Lille, 59650, Villeneuve d'Ascq, France

*** Correspondence:**
Yann Beilliard
yann.beilliard@usherbrooke.ca

## Abstract

*Neuromorphic computing based on spiking neural networks has the potential to significantly improve on-line learning capabilities and energy efficiency of artificial intelligence, specially for edge computing. Recent progress in computational neuroscience have demonstrated the importance of heterosynaptic plasticity for network activity regulation and memorization. Implementing heterosynaptic plasticity in hardware is thus highly desirable, but important materials and engineering challenges remain, calling for breakthroughs in neuromorphic devices. In this mini-review, we propose an overview of the latest advances in multi-terminal memristive devices on silicon with tunable synaptic plasticity, enabling heterosynaptic plasticity in hardware. The scalability and compatibility of the devices with industrial complementary metal oxide semiconductor (CMOS) technologies are discussed.*

## 1. Introduction

For artificial intelligence (AI) applications, traditional computer hardware based on the serial von Neumann architecture suffers from a major performance and energy-efficiency, mainly due to the massive data transfer between processing and memory units. One promising way to overcome this so-called von Neumann bottleneck is to embrace computational neuroscience and neuromorphic engineering paradigms to create parallelized spike-based computing systems (Yang et al., 2020). In that scope, emerging two-terminal resistive memories (i.e. memristors) are considered as key building blocks, owing to their controllable conductance state which can be used to encode synaptic weights (Li and Ang, 2021). The main synaptic-like behaviors reported to date include short- and long-term memory (STM/LTM), paired-pulse facilitation (PPF) and depression (PPD), spike-timing-dependent plasticity (STDP), spike-rate-dependent plasticity (SRDP) and metaplasticity (Zhu et al., 2020).

The two-terminal architecture of typical memristive devices can only authorize the emulation of homosynaptic plasticity – input-specific changes at synapses directly involved in a pre-synaptic activation. Hence, most of the studies done on memristor-based spiking neural networks (SNN) rely solely on Hebbian-type learning rules (e.g. STDP) based on homosynaptic plasticity (Chua et al., 2019). Such SNNs are prone to weight saturation (i.e. runaway dynamics) and are devoid of synaptic competition, which can weaken network representativity due to overexcitability or silencing of neurons (Moser and Moser, 1999; Chen et al., 2013). These phenomena are usually at the source of poor learning and computing performance. Recent advances in neuroscience have demonstrated that synaptic plasticity in biological neural networks is regulated not only by homosynaptic plasticity, but also by synaptic interactions (competition, cooperation) and more importantly heterosynaptic plasticity

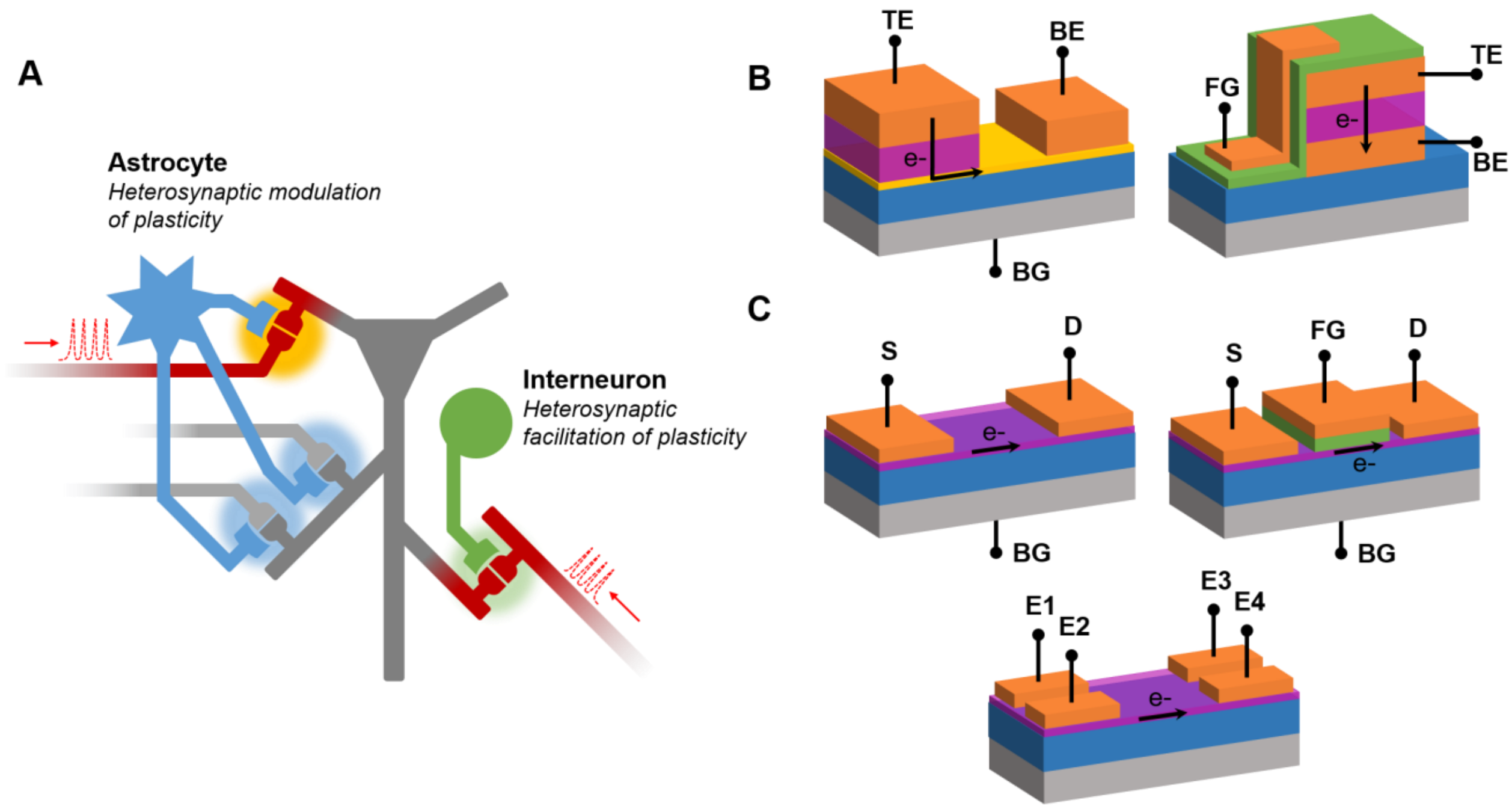

**Figure 1. (A)** Schematic representations of spiking activities between neurons, modulated by an astrocyte and an interneuron. Homosynaptic plasticity occurs at synapses directly involved in a pre-synaptic activation (red connections). An astrocyte can induce heterosynaptic plasticity modulation in distant inactive synapses (blue areas) based on local synaptic activities (yellow area). An interneuron can act as an intermediate cell responsible for heterosynaptic facilitation (green area). **(B, C)** Schematic representations of recently reported artificial synapses based on emerging multi-terminal vertical and planar memristive devices, featuring resistive switching materials (transparent purple) and electrodes (orange). **(B)** Typical vertical multi-terminal devices, where the modulation electrode is implemented either by the substrate as a back gate (BG) or a top gate (TG) separated from the switching materials by a dielectric layer (green). The top electrode (TE) and the bottom electrode (BE) are placed at different altitudes, and both the switching mechanisms and the electronic current (e-) are vertical. **(C)** Various architectures of planar multi-terminal devices where the switching mechanisms and the current electronic are parallel to the substrate. In a memtransistor architecture, two electrodes are used as source (S) and drain (D), and the switching modulation is done using a back gate and/or a front gate. Other works also involve devices with multiple electrodes fabricated on top of the switching material, allowing to implement synaptic interactions.

– modulatory input-dependent changes at inactive synapses related to strong post-synaptic activities (Bailey et al., 2000; Chistiakova et al., 2014). These complex heterosynaptic behaviors, which can be mediated by limited available energy and modulatory cells (e.g. astrocyte and interneurons, see Figure 1A), play an important role in the stabilisation and memory functions of neural systems. They are indeed responsible for the normalization of every synaptic weights connected to a neuron, meaning that the total weight remains constant. Such regulation mechanisms strongly limit the risk of synaptic weights to saturate, and thus contribute to improve the learning capability and computing performance of the network. Therefore, it appears that homosynaptic and heterosynaptic plasticity mechanisms have complementary computational properties at different spatiotemporal scales that could be beneficial for neuromorphic systems.

In recent years, the use of heterosynaptic plasticity in software SNNs has resulted in improved learning capabilities (Susi et al., 2018; Daram et al., 2020). The implementation of an astrocyte-reinforced STDP has also been reported in Intel's Loihi neuromorphic chip (Tang et al., 2019). Going forward, improving the online learning capabilities, the speed and the energy efficiency of future bio-inspired computing systems will require novel neuromorphic hardware that natively emulate both homo- and heterosynaptic plasticity mechanisms. For this reason, multi-terminal memristive devices have attracted a lot of attention lately due to their tunable and rich resistive switching dynamics at various time scales, enabled by the interplay of local modulatory voltage signals and complex nanostructures based emerging materials. In addition, multi-terminal devices offer the possibility to go beyond the simple synaptic element and to engineer multiple input/output components, thus increasing the fan-in/fan-out at the device level without requiring complex interconnections. As depicted in Figures 1B and 1C, both vertical and planar multi-terminal architectures are currently investigated with the goal to mimic complex plasticity mechanisms such as heterosynaptic plasticity and synaptic interactions between multiple electrodes.

In this mini-review, our aim is to provide a brief overview of the latest progress in multi-terminal memristive devices paving the way for the hardware implementation of both homo- and heterosynaptic plasticity. We will focus our analysis of the reported studies in the light of their compatibility with industrial complementary metal oxide semiconductor (CMOS) technologies. The compatibility of multi-terminal memristive devices with the semiconductor industry is indeed of major importance for future endeavors aiming to develop truly scalable low-power high-performance neuromorphic systems. As such, this article specifically targets inorganic solid-state memristive devices fabricated on silicon substrate having the potential to be co-integrated with CMOS circuits. This article will first outline multi-terminal devices with a vertical architecture, before reviewing planar devices. For both sub-sections, reported works include memristors and memtransistors based on metal oxides, nanoparticles, nanowires, 2D materials and heterostructures. The last part of this article will discuss future research opportunities and development directions aiming to tackle challenges related to CMOS-compatible fabrication, device performance and scalability.

## 2. Multi-terminal devices as tunable artificial synapses

This section reviews the state-of-the-art of vertical and planar memristive devices with three or more terminals enabling tunable synaptic plasticity and synaptic interactions. As shown in Figures 1B and 1C, devices are classified as vertical (planar) when the switching mechanisms and the electronic current are perpendicular (parallel) to the substrate. All devices reviewed are summarized in Table 1.

### 2.1. Vertical devices

(Yang et al., 2017) reported a three-terminal oxide-based memristive device where the conventional two-terminal synapse was based on a Ta/$TaO_x$/Pt structure, while a Pt/Ti side electrode was used as a modulatory electrode. A similar three-terminal architecture was investigated by (Herrmann et al., 2018), with the specificity of having the $SrTiO_3$-based active layer being deposited on the side of the top and bottom electrodes. However, only (Yang et al., 2017) used this architecture to modulate the migration dynamics of the oxygen vacancies in the active layer to achieve tunable LTM. Synaptic facilitation/depression was reported using modulation voltage of up to 3 V. Interestingly, a slight difference in switching dynamics was noticed depending on the sign of the modulation voltage, suggesting non-uniformity in the distribution of the enhanced electric field. This result indicates that various modulation gate geometries should be explored in future studies in order to improve the control over the switching modulation. Finally, in-memory Boolean logic and simulations of MNIST classification (machine learning) were reported. In the latter case, the modulation gate was used as a mean to dynamically change the learning rate, paving the way for both accelerated learning and higher accuracy.

Other reported works involve the use of 2D materials and heterostructures. In 2015, a three-terminal device based on graphene combining a transistor and a memristor was proposed by (Tian et al., 2015b). The active layer was made of a 5 nm thick native aluminum oxide ($AlO_x$) formed at the interface between an aluminum-based top electrode and the graphene layer used as a bottom electrode. The resistive switching behavior, attributed to the formation of a conductive filament (CF) made of oxygen vacancies in the $AlO_x$ layer, was modulated by applying a voltage on the back gate (i.e. the Si substate) ranging from -35 V to 35 V. In doing so, the resistive switching window could be tuned thanks to the band gap opening in the bilayer graphene layer. Another work published by (Tian et al., 2017) reported both excitatory and inhibitory synaptic behavior in a vertical two-terminal device where the back gate acted as a modulation terminal. The device consisted in a heterojunction formed by black phosphorus (BP) and tin selenide (SnSe) deposited by exfoliation on $SiO_2$. A phosphorus oxide ($PO_x$) was intentionally formed between the BP and $SiO_2$ layers to act as an electron trap. The interplay between the charge transfer at the BP/$PO_x$ interface and the back-gate-tunable electronic properties of the SnSe/BP heterojunction gave rise to reconfigurable synaptic characteristics capable of excitatory or inhibitory responses. Further exploration of heterostructures was done by (Huh et al., 2018), with a synaptic device consisting in a vertical Ag/$WO_{3-x}$/$WSe_2$/graphene architecture. While the $WO_{3-x}$ served as the main memristive layer, the $WSe_2$/graphene heterojunction acted as a gate-controlled Schottky barrier, so-called the “barristor”. Advanced synaptic functions were successfully emulated, including STM, LTM and PPF. The use of large modulation voltages of up to 85 V allowed to tune the switching windows as well as the programing dynamics, including transition from STP to LTP.

More recently, (Guo et al., 2020) reported a vertical Ag/$SnO_x$/SnSe/Ni memristor fabricated using a van der Waals (vdW) metal-integration approach. Although this fabrication process (Liu et al., 2019) is not yet fully scalable and compatible with conventional semiconductor technologies, it allowed the authors to fabricate the Ag top electrode without damaging the $SnO_x$-based switching junction. A 100 % fabrication yield was achieved, and the devices exhibited promising endurance and retention performance. The formation of Ag filaments in the $SnO_x$ layer was identified as the main switching mechanisms. In terms of synaptic behavior, tunable STM, STM to LTM and STDP were demonstrated with a modulation voltage of up to 60 V on the back gate. However, the duration of the programing voltage pulses was in the range of seconds, which would cause serious performance issues in the case of hardware-implemented neural networks. Finally, (Choi et al., 2021) have recently demonstrated the first crossbar integration of back-gated memristors made of $SiO_x$/graphene bi-layer. The 16×16 memristor array was used to implement non-volatile universal logic gates. Along with all the other vertical devices discussed in this section, this demonstration represents a step towards a scalable hardware implementation of neuromorphic computing leveraging tunable synaptic plasticity

## 2.2. Planar devices

### 2.2.1. Single structures

Recent advances in nanofabrication and material synthesis have allowed numerous groups to explore the use of emerging 2D materials as a resistive switching medium for planar multi-terminal synaptic devices. The atomically thin geometry of 2D materials presents multiple advantages in terms of scalability, electrical tunability and low energy consumption. In recent years, one of the most studies 2D materials for tunable neuromorphic functions is molybdenum disulfide ($MoS_2$). In monolayer $MoS_2$ with defects, sulfur vacancies can act as electron donors which can promote the migration of ions and the formation of CFs made of active metal atoms provided by electrodes (Wang et al., 2020). In addition to its atomic scalability, this transition metal dichalcogenide offers fast charge carrier dynamics (Hong et al., 2014) and tunable electronic, optoelectronic and electrochemical properties (Radisavljevic et al., 2011; Mak and Shan, 2016) which can be leveraged to implement multi-gated neuromorphic devices (John et al., 2018; Jadwiszczak et al., 2019).

For synaptic applications, Sangwan *et al*. proposed gate-tunable memtransistor devices made of polycrystalline monolayer $MoS_2$ with various configurations: two terminals and a back gate (Sangwan et al., 2015), three terminals including a top gate (Sangwan et al., 2018b) and six terminals with a back gate (Sangwan et al., 2018a). With the use of the back gate, tunable resistive switching windows were observed during DC voltage sweeps using a modulation voltage ranging from -50 V to 55 V. The top gate architecture used in (Sangwan et al., 2018b) allowed to divide by 10 the required voltage to observe modulation. Low gate voltage in the range of a few volts were also observed by (Wang et al., 2019) in the case of $MoS_2$-based device with a top gate and $HfO_2$ as gate insulator. Regarding the six-terminal devices (Sangwan et al., 2018a), LTM and STDP were demonstrated using voltage pulses in the range of 40 V. Interestingly, synaptic interactions were achieved by increasing (decreasing) the conductance of the $MoS_2$ using the most distant terminals of the 6-terminal configuration, which induced an increase (decrease) of conductance between the inner terminals. This type of synaptic modulation could be used to implement a form of cooperative heterosynaptic plasticity.

Another group have successfully fabricated a dual-gated device based on exfoliated $MoS_2$ (He et al., 2020a). Using h-BN as a gate dielectric, versatile synaptic behavior was shown including tunable resistive switching window and LTM. Authors have also reported simulation results indicating that a better MNIST classification accuracy could be obtained by using a modulation voltage comprised between -5 and 5 V. (Zhu et al., 2019) have explored the use of exfoliated $Li_xMoS_2$ to fabricate a five-terminal planar device. The switching mechanisms was attributed to the transition of $MoS_2$ between the 2H (semiconductor) and 1T′ (metal) phases induced by an increase/decrease in the local $Li^+$ ion concentration. Three- and five-terminal configurations were used to demonstrate synaptic competition and cooperation. The competition was driven by the limited amount of $Li^+$ ions in the $Li_xMoS_2$ film between two connected synapses. For cooperation, the five-terminal architecture allowed to observe that the potentiation (depression) of one synapse induced the potentiation (depression) of the four others.

Other 2D materials than $MoS_2$ were also explored for planar devices. (Yang et al., 2019) have used layered GaSe nanosheets to fabricate a three-terminal synaptic device, exhibiting back-gate-tunable non-volatile switching using a low electric field of $3.3 \times 10^2$ V $cm^{-1}$. The latter was attributed to the low migration energy of the intrinsic Ga vacancy in p-type GaSe. However, the relatively high switching voltage (~V) and current level (~mA) suggest that integration optimisations remain to be done to employ this type of device in low-power applications. Another gate-tunable synaptic device was reported by (Zhao et al., 2020) where the switching junction was made of non-layered $In_2S_3$ flakes.

The latter were deposited by chemical vapor deposition (CVD), which required a thermal treatment of 980°C for 20 minutes. One can note that such temperature is not compatible with industrial back-end-of-line (BEOL) processes limited to 400 °C. Nevertheless, both tunable switching window and LTM were demonstrated with either optical or electrical modulation using the substrate as a back gate. Interestingly, depending on the back gate polarity, the device was either potentiated or depressed without changing the input pulse polarity. This controllable excitatory or inhibitory response is similar to what was observed in vertical SnSe-based devices by (Tian et al., 2017).

More recently, memristive devices based on ferroelectric switching α-$In_2Se_3$ were reported by (Xue et al., 2021). A six-terminal architecture was proposed, enabling multiple biasing configurations for both non-volatile DC switching and pulse programing. The required pulse duration was in the range of seconds, which would be redhibitory for a large-scale neuromorphic system. Nevertheless, synaptic interaction was observed as the modification of conductance between a pair of electrodes induced a change of conductance between remote pairs of electrodes. A three-terminal configuration was used to demonstrate synaptic change induced by the third electrode, as well as synaptic cooperation and competition by sending voltage pulses in 2 independent electrodes. The possibility to use multiple electrodes to directly change the conductance of the device is complementary to the gate-assisted tuning of plasticity. In the future, combining these two mechanisms would grant even more flexibility and control in the switching dynamics, enabling richer synaptic functions in hardware.

Only a few studies have reported heterosynaptic plasticity on planar devices without using 2D materials. (Nagata et al., 2019) have employed rutile $TiO_{2-x}$ single crystal for thin film memristive planar devices with a four-terminal architecture. The multi-terminal device relied on field-induced oxygen vacancies migration to implement gate-tunable gradual LTM with a very low cycle-to-cycle variability. The use of 2 electrodes for pulse programming and the other 2 electrodes as modulation gates allowed to improve the endurance and to tune the switching dynamics. Although these results are promising, the long voltage pulse in the range of seconds, the low resistance in the range of a hundred Ohms and the high temperature treatment of 700 °C required to fabricate the $TiO_{2-x}$ layer currently prevent the use of such devices in an actual large-scale system. Another interesting approach explored by (Yang et al., 2015) consisted in a device where the switching layer was made of Ag nanoclusters deposited on a dielectric material. A four-terminal architecture demonstrated the physical evolution of the CF formed from the nanoclusters depending on the direction of the electric field between the Pt electrodes. Furthermore, a three-terminal version of the device where the third terminal was used as a modulation gate allowed to implement heterosynaptic facilitation and depression. However, the combination of ultra-low current levels (~pA) and high switching voltage (~10 V) calls for further development to render such devices compatible with operation conditions of CMOS technologies.

### 2.2.2. Heterostructures

The use of heterostructures combining dielectric thin films and 2D materials has been explored by several groups to implement complex synaptic behaviors. Building upon their work in (Tian et al., 2015b), (Tian et al., 2015a) have investigated a Al/$AlO_x$/graphene architecture. The conductance of the device could be modified by trapping carriers from the graphene inside the $AlO_x$ layer by applying voltage pulses on the top gate. The modulation of STM, LTM and STDP was demonstrated by applying a constant voltage on the Si substrate used as back gate. Interestingly, the switching dynamics could be reversed depending on the polarity and amplitude of the back gate voltage. As in (Tian et al., 2017), this phenomenon enabled the use of such devices as either excitatory or inhibitory synapses. (Chen et al., 2019) also used graphene in a four-terminal memristive vdW heterostructure combining $MoS_2$ and h-BN exfoliated on a $Al_2O_3$ thin film. Authors managed to implement synaptic plasticity at various time scales using a dual-gate approach. In this case, the top gate was used for both resistive switching and synaptic plasticity modulation induced by voltage pulses coming from the back gate. (He et al., 2020b) have also developed a dual-gated device based on a lateral heterostructure composed of 2D $WSe_2$ and $WO_3$. They successfully implemented tunable switching window, STM and LTM. While the post synaptic current was driven by the so-called source and drain electrodes, the synaptic plasticity was modulated by both electrical and optical signals applied to the back gate and top side of the device, respectively. The switching mechanism was attributed to the injection/extraction of protons in the $WO_3$ layer, which can be electrostatically tuned. The response to visible light of the device was explained by the creation of photogenerated holes in the $WSe_2$ 2D material upon illumination. The use of multiple modulation gates using different physical mechanisms offers a higher degree of control over the synaptic plasticity, which could prove useful for neuromorphic optoelectronic systems. Finally, another approach based on Ag nanowires with an insulating shell made of polyvinylpyrrolidone (PVP) was explored by (Milano et al., 2020) to fabricate a memristive network exhibiting advanced synaptic plasticity behavior. Using multiple terminals connected on the edge of the nanowire network, authors demonstrated that the device could emulate multiple synaptic connections and pathways showing resistive switching, STM, PPF and heterosynaptic facilitation.

**Table 1**. Summary of reported multi-terminal vertical and planar memristive devices.

| **Device type** | **Material stacks** | **Number of physical terminals** | **Range of switching voltage (V)** | **Max. current (A)** | **Range of Modulation Voltage (V)** | **Synaptic functions demonstrated** | **Ref.** |
|---|---|---|---|---|---|---|---|
| Vertical | $Ta/TaO_x/Pt$ | 3 | DC: [-1.5; 1.5]<br>Pulse: [-1.55;1] | $10^{-3}$ | DC: [-3;3]<br>Pulse: [-3;3] | Tunable LTM, tunable learning rate in MNIST classification (simulations) | (Yang et al., 2017) |
| | $Al/AlO_x$/graphene/ | 2 + back gate | DC: [-3;4]<br>Pulse: n/a | $10^{-4}$ | DC: [-35;35]<br>Pulse: n/a | Tunable switching window | (Tian et al., 2015b) |
| | Au/SnSe/BP/Au | 2 + back gate | DC: n/a<br>Pulse: [-20; 20] | $10^{-6}$ | DC: n/a<br>Pulse: [-5;3] | Tunable STM, LTM, STDP excitatory and inhibitory response | (Tian et al., 2017) |
| | $Ag/WO_{3-x}/WSe_2$/graphene | 2 + back gate | DC: [-0.7;0.3]<br>Pulse: [-4;0.5] | $10^{-7}$ | DC: [-85; 45]<br>Pulse: [-40; 0] | Tunable switching window, tunable STM to LTM, PPF | (Huh et al., 2018) |
| | $Ag/SnO_x$/SnSe/Ni | 2 + back gate | DC: [-2; 2]<br>Pulse: [-1; 1] | $10^{-6}$ | DC: [-60;60]<br>Pulse: [-60;60] | Tunable switching window, tunable STM to LTM, tunable STDP | (Guo et al., 2020) |
| | $Pd/SiO_x$/graphene/Pd<br>(crossbar) | 2 + back gate | DC: [0;12]<br>Pulse: [4;15] | $10^{-4}$ | DC: [0;20]<br>Pulse: [0;20] | Tunable unipolar switching window | (Choi et al., 2021) |
| Planar | $Au/MoS_2/Au$ | 2 + back gate | DC: [-40;35]<br>Pulse: [-20;20] | $10^{-9}$-$10^{-5}$ | DC: [0;25]<br>Pulse: n/a | Tunable switching window | (Jadwiszczak et al., 2019) |

| | | | | | | |
|---|---|---|---|---|---|---|
| Au/$MoS_2$/Au | 3 | DC: [-12;12]<br>Pulse: [-10;10] | $10^{-6}$ | DC: [-6;4]<br>Pulse: n/a | Tunable switching window | (Wang et al., 2019) |
| Au/$MoS_2$/Au | 2 + back gate | DC: [-10; 55]<br>Pulse: n/a | $10^{-6}$ | DC: [-40;50]<br>Pulse: n/a | Tunable switching window | (Sangwan et al., 2015) |
| Au/$MoS_2$/Au<br>(top gate configuration) | 3 | DC: [0;15]<br>Pulse: n/a | $10^{-8}$ | DC: 5<br>Pulse: n/a | Tunable switching window | (Sangwan et al., 2018b) |
| Au/$MoS_2$/Au | 6 | DC: [-80;80]<br>Pulse: [-40;40] | $10^{-4}$ | DC: [-40;40]<br>Pulse: n/a | Tunable switching window, synaptic interaction | (Sangwan et al., 2018a) |
| Au/$MoS_2$/Au | 3 + back gate | DC: [-13;13]<br>Pulse: [-18;22] | $10^{-6}$ | DC: [-13;13]<br>Pulse: [-10;10] | Tunable switching window, tunable LTM, tunable linearity, tunable learning rate (simulation) | (He et al., 2020a) |
| Au/$Li_xMoS_2$/Au | 5 | DC: [-6; 6]<br>Pulse: [-6;6] | $10^{-5}$ | DC: n/a<br>Pulse: n/a | LTM, synaptic competition and cooperation | (Zhu et al., 2019) |
| Ag/GaSe/Ag | 2 + back gate | DC: [-2;2]<br>Pulse: n/a | $10^{-3}$ | DC: [-20;20]<br>Pulse: n/a | Tunable switching window | (Yang et al., 2019) |
| Au/$In_2S_3$/Au | 2 + back gate | DC: [-20;20]<br>Pulse: [-6;10] | $10^{-6}$ | DC: [-40;40]<br>Pulse: [-10;60] | Tunable switching window (optical, electrical), tunable LTM, excitatory and inhibitory response | (Zhao et al., 2020) |
| Au/α-$In_2Se_3$/Au | 6 | DC: [-5;5]<br>Pulse: [-4;4] | $10^{-10}$ | DC: n/a<br>Pulse: [-4; 4] | Resistive switching (DC), heterosynaptic plasticity by multi-terminal programing, synaptic cooperation and competition | (Xue et al., 2021) |
| Pt/$TiO_2$/Pt | 4 | DC: [-6;6]<br>Pulse: [-4;4] | $10^{-2}$ | DC: n/a<br>Pulse: [0; 4] | Multilevel resistive switching (DC), tunable LTM | (Nagata et al., 2019) |
| Pt/Ag/Pt | 3 | DC: [8; 35]<br>Pulse: 10 | $10^{-10}$ | DC: n/a<br>Pulse: [-8;8] | Physical evolution of the Ag conductive filament, tunable LTM | (Yang et al., 2015) |
| Al/$AlO_x$/graphene | 3 + back gate | DC: [-2; 2] | $10^{-6}$ | DC: n/a | Tunable STM, LTM, STDP excitatory and | (Tian et al., 2015a) |
| | | Pulse: [-2; 2] | | Pulse: [-40;40] | inhibitory response | |
| Au/$MoS_2$/Au on $Al_2O_3$ | 3 + back gate | DC: [-10; 10]<br>Pulse: -8 | $10^{-6}$ | DC: n/a<br>Pulse: [-1,1] | Resistive switching, STM, LTM, PPF, tunable LTM | (Chen et al., 2019) |
| Pd/$WSe_2$/$WO_3$/Pd | 3 + back gate | DC: [-5; 5]<br>Pulse: [-2; 3] | $10^{-7}$ | DC: [-40; 40]<br>Pulse: [-2; 3] | Tunable switching window, tunable STM and LTM using voltage and light | (He et al., 2020b) |
| Au/PVP/Ag/Au (nanowire) | 5 | DC: [-6; 6]<br>Pulse: [1; 15] | $10^{-4}$ | DC: n/a<br>Pulse: n/a | STM, PPF, synaptic interaction, heterosynaptic facilitation | (Milano et al., 2020) |

## 3. Discussion and perspectives

The growing necessity to implement advanced synaptic functions in neuromorphic hardware has pushed several groups to explore novel multi-terminal memristive devices based on metal oxides, nanoparticles, nanowires and 2D materials. Rich and tunable synaptic-like behaviors have been demonstrated, including short- and long-term heterosynaptic plasticity as well as synaptic interactions. Lately, 2D materials such as graphene, SnSe, $MoS_2$, $WSe_2$, GaSe and $In_2Se_3$ have emerged as promising candidates owing to their atomic-level thickness and tunable electronic properties that offer lower energy consumption and improved synaptic emulation. Although these demonstrations contribute to bridging the gap between software and hardware in the field of neuromorphic engineering, major challenges remain in terms of materials, device operation conditions, design and integration with CMOS technologies.

Materials and fabrication processes compatible with CMOS technologies have to be investigated. Outside of memristive devices based on oxide thin films (Yang et al., 2017; Nagata et al., 2019), most demonstrations involve 2D materials which rely on fabrication processes that are not scalable such as exfoliation process (Yang et al., 2019; Zhu et al., 2019) or patterning with focused ion beam (Jadwiszczak et al., 2019). Some works employed wafer-level chemical vapor deposition (CVD), but the required thermal treatment is way above 400 °C (Sangwan et al., 2018a; Wang et al., 2019; Zhao et al., 2020) which makes this approach still incompatible with CMOS processes. Another important aspect in that regards is the overabundant use of Au and Pt as electrodes. The use of inert materials readily available in foundries (e.g. TiN, W) should be investigated, as well as their impact on resistive switching mechanisms.

Regarding switching operations, reported devices relying on a back gate to modulate the switching operations necessitate gate voltage in the range of tens of volts. Planar devices based on 2D materials can also exhibit very high switching voltages, with values ranging from 20 V to 80 V (see Table 1). For tunable memristive devices to be co-integrated with CMOS technologies, breakthroughs in fabrication and materials engineering remain to be made for the operation conditions to be in the range of a few volts. For example, as shown in some studies (Yang et al., 2017; Sangwan et al., 2018b), the integration of the modulation electrode as a top/side gate instead of a back gate can reduce the required voltage by an order of magnitude. Advanced designs such as gate-all-around architecture could potentially improve further the control on synaptic modulation.

In terms of scalable integration, one important milestone is to achieve the integration of synaptic devices into a fully connected network. In that aspect, further work has to be done to convert current planar devices into vertical multi-terminal memories. The vertical architecture indeed allows the fabrication of crosspoint structures, which provide the highest integration density when interconnected into a crossbar array. Furthermore, vertical integration opens new perspectives for monolithic 3D integration. In fact, conventional 2D integration requires complex interconnection layouts to mimic the parallel structure of the brain. Integrating multiple input/output at the device level and taking advantage of 3D interconnects with both lateral and vertical access lines (Xue et al., 2021) could help relaxing the important limitation of todays hardware for neuromorphic engineering. In the meantime, only one group has reported the fabrication of crossbar arrays of multi-terminal memories (Choi et al., 2021). In addition to device density, this approach allows to scale down the switching material thickness which should help to optimize the operation conditions, in particular the programing pulse duration which can be in the range of seconds for some planar devices (Nagata et al., 2019; Xue et al., 2021). Therefore, extensive studies and demonstrations of scalable multi-terminal devices with a crosspoint structure represent promising research opportunities.

Some groups have started to investigate the benefits of gate-tunable plasticity in neural network simulations, reporting the ability to optimise the learning rate and improve MNIST classification accuracy (Yang et al., 2017; He et al., 2020a). Going further, efforts should also be focused on hardware implementation of heterosynaptic plasticity at the network level, as well as software-hardware co-design aiming towards weight normalization or heterosynaptic metaplasticity (Hulme et al., 2014). Only hardware-based demonstrations at the network level will allow to evaluate if these novel learning capabilities enabled by advanced memristive devices are worth potential increases in energy consumption or design requirements in neuromorphic systems.

## 4. Conflict of Interest

The authors declare that the research was conducted in the absence of any commercial or financial relationships that could be construed as a potential conflict of interest.

## 5. Author Contributions

YB and FA conceived the review topic. YB wrote the paper. All authors contributed to the article and approved the submitted version.

## 6. Funding

We acknowledge financial supports from the EU: ERC-2017-COG project IONOS (# GA 773228), the Natural Sciences and Engineering Research Council of Canada (NSERC) HIDATA project 506289-2017 and CHIST-ERA UNICO project.